\documentclass[submission,copyright,creativecommons]{eptcs}
\providecommand{\event}{VPT 2021} 
\usepackage{breakurl}             
\usepackage{underscore}           
\usepackage{tikz}
\usetikzlibrary{shapes,arrows,automata,positioning,quotes,backgrounds,decorations.text,decorations.pathmorphing,calc}
\usepackage{bold-extra}

\usepackage{booktabs}   
\usepackage{multirow}

\usepackage{listings}

\usepackage{graphicx}

\usepackage{subcaption}
\usepackage{proof}

\usepackage{amsmath}

\usepackage{xcolor}

\usepackage{xspace}

\usepackage{wrapfig}

\usepackage{array}
\newcolumntype{e}[1]{>{\centering\let\newline\\\arraybackslash\hspace{0pt}}m{#1}}

\lstdefinelanguage{ocanren}{
keywords={run, conde, fresh, let, in, match, with, when, class, type,
object, method, of, rec, repeat, until, while, do, done, as, val, inherit,
new, module, sig, deriving, datatype, struct, if, then, else, open, private, virtual, include, success, failure,
true, false, none, some, integer, zero, succ, prod, num, sum},
sensitive=true,
commentstyle=\small\itshape\ttfamily,
keywordstyle=\textbf,
identifierstyle=\ttfamily,
basewidth={0.5em,0.5em},
columns=fixed,
mathescape=true,
fontadjust=true,
literate={fun}{{$\lambda$}}1 {->}{{$\to$}}3 {===}{{$\equiv$}}1 {=/=}{{$\not\equiv$}}1 {|>}{{$\triangleright$}}3 {\\/}{{$\vee$}}2 {/\\}{{$\wedge$}}2 {^}{{$\uparrow$}}1,
morecomment=[s]{(*}{*)}
}

\newcommand{\mk}{\textsc{miniKanren}\xspace}
\newcommand{\muk}{\textsc{microKanren}\xspace}
\newcommand{\oc}{\textsc{OCanren}\xspace}
\newcommand{\ocaml}{\textsc{OCaml}\xspace}
\newcommand{\haskell}{\textsc{Haskell}\xspace}
\newcommand{\pro}{\textsc{Prolog}\xspace}
\newcommand{\scheme}{\textsc{Scheme}\xspace}

\newcommand{\ecce}{\textsc{ECCE}\xspace}

\newcommand{\conj}{$\wedge$\xspace}
\newcommand{\rel}[2]{\texttt{#1}$^o$ #2}
\newcommand{\subst}[1]{$\langle$#1$\rangle$}

\tikzstyle{processTree} = [
  ->,
  sibling distance=15em,
  scale=0.6,
  node distance=2.5cm,
  every node/.style = {
    shape=rectangle,
    rounded corners=0.05cm,
    draw,
    align=center,
    minimum size=5mm,
    scale=0.6,},
  ]

\tikzstyle{program} = [
  draw=black,
  thick,
  rectangle,
  rounded corners=1pt,
  inner sep=5pt,
  inner ysep=5pt
  ]

\tikzstyle{goal} = [
  draw=black,
  rectangle,
  rounded corners=1pt,
  inner ysep=0pt,
  ]

\tikzstyle{input} = [
  draw=none,
  rectangle,
  rounded corners=1pt,
  inner sep=2pt,
  inner ysep=2pt,
  fill=green!10,
  minimum height=5mm
  ]

\tikzstyle{transparent} = [
  draw=none,
  inner ysep=3pt
  ]

\title{An Empirical Study of Partial Deduction for \mk}

\author{Ekaterina Verbitskaia
\institute{JetBrains Research\\
Saint Petersburg, Russia}
\email{kajigor@gmail.com}
\and
Daniil Berezun
\institute{Saint Petersburg State University}
\institute{JetBrains Research\\
Saint Petersburg, Russia}
\email{d.berezun@2009.spbu.ru}
\and
Dmitry Boulytchev
\institute{Saint Petersburg State University}
\institute{JetBrains Research\\
Saint Petersburg, Russia}
\email{dboulytchev@math.spbu.ru}
}
\def\titlerunning{An Empirical Study of Partial Deduction for \mk}
\def\authorrunning{E. Verbitskaia, D. Berezun \& D. Boulytchev}
\begin{document}

\maketitle

\begin{abstract}
  We study conjunctive partial deduction, an advanced specialization technique aimed at improving the performance of logic programs, in the context of relational programming language \mk. We identify a number of issues, caused by  \mk peculiarities, and describe a novel approach to specialization based on partial deduction and supercompilation. The results of the evaluation demonstrate successful specialization of relational interpreters. Although the project is at an early stage, we consider it as the first step towards an efficient optimization framework for  \mk.
\end{abstract}

\section{Introduction}
\label{intro}

A family of embedded domain-specific languages \mk\footnote{\mk language web site: \url{http://minikanren.org}. Access date: 28.02.2021} implement relational programming~---~a paradigm closely related to pure logic programming.
The minimal core of the language, also known as \muk, can be implemented in as little as 39 lines of \scheme~\cite{friedmanmukanren}.
An introduction to the language and some of its extensions in a series of examples can be found in the book~\cite{TheReasonedSchemer}.
The formal certified semantics for \mk is described in~\cite{rozplokhas2020certified}.

Relational programming is a paradigm based on the idea of describing programs as relations.
The core feature of relational programming is the ability to run a program in various directions by executing goals with free variables.
The distribution of free variable occurrences determines the direction of relational search.
For example, having specified a relation for adding two numbers, one can also compute the subtraction of two numbers or find all pairs of numbers which can be summed up to get the given one.
One of the most prominent applications of relational programming amounts to implementing interpreters as relations.
By running a relational interpreter for some language \emph{backwards} one can do program synthesis.
In general, it is possible to create a solver from a recognizer by translating it into \mk and running it in the appropriate direction~\cite{lozov2019relational}.

The search employed in \mk is complete which means that every answer will be found, although it may take a long time.
The promise of \mk falls short when speaking of performance.
The execution time of a program in \mk is highly unpredictable and varies greatly for various directions.
What is even worse, it depends on the order of the relation calls within a program.
One order can be good for one direction, but slow down the computation dramatically in the other direction.

Partial evaluation~\cite{jonesbook} is a technique for specialization, i.e. improving the performance of a program given some information about it beforehand.
It may either be a known value of some argument, its structure (e.g. the length of an input list) or, in case of a relational program, the direction in which the relation is intended to be run.
An earlier paper~\cite{lozov2019relational} has shown that \emph{conjunctive partial deduction}~\cite{de1999conjunctive} can sometimes improve the performance of \mk programs.
Depending on the particular \emph{control} decisions, it may also not affect the execution time of a program or even make it slower.

Control issues in partial deduction of the logic programming language \pro have been studied before~\cite{leuschel2002logic}.
Under the left-to-right evaluation strategy of \pro, atoms in the right-hand side of a clause cannot be arbitrarily reordered without changing the observable behavior of a program.
In contrast, due to the completeness of the search, \mk is less sensitive to the order of conjuncts: no answers can be added or lost, and the only difference caused by the order of conjuncts may be the divergence/convergence of the search in the case when all answers are found.
This opens yet another possibility for optimization, not taken into account by the approaches initially developed in the context of conventional logic programming.

In this paper we make the following contributions.
We study issues which conjunctive partial deduction faces being applied for \mk.
We also describe a novel approach to partial deduction for relational programming, \emph{conservative partial deduction}.
We implemented this approach and compared it with the existing specialization system (\ecce) for several programs.
We report here the results of the comparison and discuss why some \mk programs run slower after specialization.

\section{Background}
\label{background}

In this section we provide some background on relational programming and relational interpreters.

\subsection{\mk}
\label{mkIntro}

\begin{figure*}[t]
  \[
  \begin{array}{cccll}
    &\mathcal{T} & = & \mathcal{X} \cup \{C_i^{k_i} (t_1, \dots, t_{k_i}) \mid t_j\in\mathcal{T}\} & \mbox{terms over the set of variables $\mathcal{X}$} \\
    &\mathcal{G} & = & \mathcal{T}\equiv\mathcal{T}   &  \mbox{unification} \\
    &            &   & \mathcal{G}\wedge\mathcal{G}     & \mbox{conjunction} \\
    &            &   & \mathcal{G}\vee\mathcal{G}       &\mbox{disjunction} \\
    &            &   & \mbox{\lstinline|fresh|}\;\mathcal{X}\;.\;\mathcal{G} & \mbox{fresh variable introduction} \\
    &            &   & R_i^{k_i} (t_1,\dots,t_{k_i}),\;t_j\in\mathcal{T} & \mbox{relational symbol invocation} \\
    &\mathcal{S} & = & \{R_i^{k_i} = \lambda\;x_1^i\dots x_{k_i}^i\,.\, g_i;\}\; g & \mbox{specification}
  \end{array}
  \]
  \caption{The syntax of the source language}
  \label{syntax}
  \end{figure*}

This paper considers the minimal relational core of the \mk language.
The syntax of the language is presented in Fig.~\ref{syntax}.
A specification of the \mk program consists of a set of \emph{relation definitions} accompanied by a top-level \emph{goal} which plays the role of a query.
Goals, being the central syntactic category of the language, can take the form of either term \emph{unification}, \emph{conjunction} or \emph{disjunction} of goals, a \emph{fresh} syntactic variable introduction, or a relation \emph{call}.
We consider the alphabet of constructors $\{C^{k_i}_i\}$ and relational symbols $\{R^{k_i}_i\}$ to be predefined and accompanied with their arities.

The formal semantics of the language is best described in~\cite{rozplokhas2020certified}.
Here we only briefly introduce the semantics.
A stream of substitutions for free variables within the query goal is computed during the execution of a \mk program.
Depending on the kind of the goal, one of the following situations is possible.

\begin{enumerate}
  \item Term unification $t_1 \equiv t_2$ computes the most-general unification in the context of the current substitution. If it succeeds, the unifier is added into the current substitution and then it is returned as a singleton stream. Otherwise, an empty stream is returned.
  \item Introduction of a fresh variable $fresh \ x. g$ allocates a new \emph{semantic} variable, substitutes it for all fresh occurrences of $x$ within $g$, then evaluates the goal.
  \item An execution of a relational call $R^{k_i}_i(t_1, \dots, t_{k_i})$ is done by first substituting the terms $t_j$ for the respective formal parameters and then running the resulting goal.
  \item When executing a conjunction $g_1 \wedge g_2$, first the goal $g_1$ is run in the context of the current substitution which results in the stream of substitutions, in each of which $g_2$ is run. The resulting stream of streams is then concatenated.
  \item Disjunction $g_1 \vee g_2$ applies both goals to the current substitution, incrementally switching evaluation steps between the subgoals until all results (if any) are found.
\end{enumerate}

Consider the relation \lstinline{add$^o$} in Listing~\ref{eval:arith}.
It defines the relation between three Peano numbers $x$, $y$ and $z$, such that $x + y = z$, using the \oc language\footnote{\oc: statically typed \mk embedding in \ocaml. The repository of the project: \url{https://github.com/JetBrains-Research/OCanren}. Access date: 28.02.2021}.
The keyword \lstinline{conde} provides syntactic sugar for a disjunction, while \lstinline{zero} and \lstinline{succ} are constructors.
The query \lstinline{fresh (z) (add$^o$ (succ zero) (succ zero) z)} results in the only substitution \lstinline{[z $\mapsto$ succ (succ zero)]}, while the query \lstinline{fresh (x y) (add$^o$ x y (succ (succ zero)))} executes to three valid substitutions: \lstinline{[x $\mapsto$ zero, y $\mapsto$ succ (succ zero)]}, \lstinline{[x $\mapsto$ succ zero,} \lstinline{y $\mapsto$ succ zero]}, \lstinline{[x $\mapsto$ succ (succ zero), y $\mapsto$ zero]}.

The \emph{interleaving} search~\cite{10.1145/1090189.1086390} is at the core of \mk.
It evaluates disjuncts incrementally, passing control from one to the other.
This search strategy is what makes the search in \mk complete.
It also allows for reordering of both disjuncts and conjuncts within a goal which may improve the efficiency of a program.
This reordering generally leads to the reordering of the answers computed by a \mk program.
The denotational semantics of \mk ignores the order of the answers because the search is complete and thus all possible answers will be found eventually.

\begin{figure*}[!t]
  \centering
  \begin{minipage}{0.68\textwidth}
    \begin{lstlisting}[label={eval:arith}, caption={Evaluator of arithmetic expressions}, captionpos=b, frame=tb, escapeinside={(*}{*)}]
  let rec add$^o$ x y z = conde [
      (x === zero /\ y === z);
      (fresh (p) (x === succ p /\ add$^o$ p (succ y) z) ) ]

  let rec eval$^o$ fm res = conde [fresh (x y xr yr) (
      (fm === (*\textbf{num}*) res);
      (eval$^o$ x xr /\  eval$^o$ y yr /\
        conde [
          (fm === (*\textbf{sum}*) x y /\ add$^o$ xr yr res);
          (fm === prod x y /\ $\dots$);
          $\dots$ ] )
    \end{lstlisting}
  \end{minipage}
\end{figure*}

\subsection{Relational Interpreters}
\label{relinterp}

The kind of relational programs most interesting to us is relational interpreters.
They may be used to solve complex problems such as generating quines~\cite{byrd2012minikanren} or to solve search problems by only implementing programs which check that a solution is correct~\cite{lozov2019relational}.
The latter application is the focus of our research project thus we provide a brief description of it.

Search problems are notoriously complicated.
In fact, they are much more complex than verification~---~checking that some candidate solution is indeed a solution.
The ability of \mk programs to be evaluated in different directions along with the complete semantics of the language allows for automatic generation of a solver from a verifier using relational conversion~\cite{lozov2017typed}.
Unfortunately, generated relational interpreters are often inefficient, since the conversion introduces a lot of extra unifications and boilerplate.
This kind of inefficiency is a prime candidate for specialization.

Consider the relational interpreter \lstinline{eval$^o$ fm res} in Listing~\ref{eval:arith}.
It evaluates an arithmetic expression \lstinline{fm} which can take the form of a number (\lstinline{num res}) or a binary expression such as the \lstinline{sum x y} or \lstinline{prod x y}.
Running the interpreter backwards synthesizes expressions which evaluate to the given number. For example one possible answer to the query \lstinline{eval$^o$ fm (succ (succ zero))} is \lstinline{sum (num (succ zero)) (sum (num zero) (num (succ zero)))}.

\section{Related Work}

Specialization is an attractive technique aimed to improve the performance of a program making use of its static properties such as known arguments or its environment.
Specialization is studied for functional, imperative, and logic programing and comes in different forms: partial evaluation~\cite{jonesbook} and partial deduction~\cite{lloyd1991partial}, supercompilation~\cite{turchin1986concept, soerensen1996positive}, distillation~\cite{hamilton2007distillation}, and many others.

The heart of supercompilation-based techniques is \emph{driving}~---~a symbolic execution of a program through all possible execution paths.
The result of driving is a possibly infinite \emph{process tree} where nodes correspond to \emph{configurations} which represent computation states.
For example, in the case of pure functional programming languages, the computational state might be a term.
Each path in the tree corresponds to some concrete program execution.
The two main sources for supercompilation optimizations are aggressive information propagation about variables' values, equalities and disequalities, and precomputing of deterministic semantic evaluation steps.
The latter process, also known as \emph{deforestation}~\cite{deforestation}, means  combining of consecutive process tree nodes with no branching.
Of course, the process tree can contain infinite branches.
\emph{Whistles} --- heuristics to identify possibly infinite branches --- are used to ensure supercompilation termination.
If a whistle signals during the construction of some branch, then something should be done to ensure termination.
The most common approaches are either to stop driving the infinite branch completely (no specialization is done in this case and the source code is blindly copied into the residual program) or to fold the process tree to a \emph{process graph}.
When the process graph is constructed, the resulting, or \emph{residual}, program can be extracted from the graph by the process called \emph{residualization}.
The main instrument to perform folding is some form of \emph{generalization}.
Generalization, abstracting away some computed data about the current term, makes folding possible.
For example, one source of infinite branches is consecutive recursive calls to the same function with an accumulating parameter: by unfolding such a call further one can only increase the term size which leads to nontermination.
The accumulating parameter can be removed by replacing the call with its generalization.
There are several ways to ensure correctness and termination of a program transformer~\cite{sorensen1998convergence}, most-specific generalization
(anti-unification) and \emph{homeomorphic embedding}~\cite{Higman52,Kruskal60} as a
whistle being common.

While supercompilation generally improves the behaviour of input programs and distillation can even provide superlinear speedup, there are no ways to predict the effect of specialization on a given program in general.
What is worse, the efficiency of a residual program from the target language evaluator point of view is rarely considered in the literature.
The main optimization source is computing in advance all possible intermediate and statically-known semantics steps at program transformation-time.
Other criteria, like the size of the generated program or possible optimizations and execution cost of different language constructions by the target language evaluator, are usually out of consideration~\cite{jonesbook}.
Partial evaluation in logic programming should be done with care to not interfere with the compiler optimizations~\cite{venken1988partial}.
It is also known that supercompilation may adversely affect GHC optimizations making standalone compilation more powerful~\cite{SCBE,TCES} and cause code explosion~\cite{SCHC}.
Moreover, it may be hard to predict the real speedup of any given program using concrete benchmarks even disregarding the problems above because of the complexity of the transformation algorithm.
The worst-case for partial evaluation is when all static variables are used in a dynamic context, and there is some advice on how to implement a partial evaluator as well as a target program so that specialization indeed improves its performance~\cite{jonesbook,bulyonkov84}.
There is a lack of research in determining the classes of programs which transformers would definitely speed~up.

Conjunctive partial deduction~\cite{de1999conjunctive} makes an effort to provide reasonable control for the left-to-right evaluation strategy of \pro.
CPD constructs a tree which models goal evaluation and is similar to an SLDNF tree; then a residual program is generated from the tree.
Partial deduction itself resembles driving in supercompilation~\cite{gluck1994partial}.
The specialization is done in two levels of control: the local control determines the shape of the residual programs, while the global control ensures that every relation which can be called in the residual program is defined.
The leaves of local control trees become nodes of the global control tree.
CPD analyses these nodes at the global level and runs local control for all those nodes which are new.

At the local level, CPD examines a conjunction of atoms by considering each atom one-by-one from left to right.
An atom is \emph{unfolded} if it is deemed safe, i.e. a whistle based on homeomorphic embedding does not signal for the atom.
When an atom is unfolded, a clause whose head can be unified with the atom is found, and a new node is added into the tree where the atom in the conjunction is replaced with the body of that clause.
If there is more than one suitable head, then several branches are added into the tree which corresponds to the disjunction in the residualized program.
An adaptation of CPD for the \mk programming language is described in~\cite{lozov2019relational}.

\ecce partial deduction system~\cite{leuschel1997ecce, LeuschelEVCF06} is the most mature implementation of CPD for \pro.
\ecce provides various implementations of both local and global control as well as several degrees of post-processing.
Unfortunately there is no automatic procedure to choose what control setting is likely to improve input programs the most.
The choice of the proper control is left to the user.

An empirical study has shown that the most well-behaved strategy of local control in CPD for \pro is \emph{deterministic unfolding}~\cite{leuschel1997advanced}.
An atom is unfolded only if precisely one suitable clause head exists for it with the one exception: it is allowed to unfold an atom non-deterministically once for each local control tree.
This means that if a non-deterministic atom is the leftmost one within a conjunction, it is most likely to be unfolded, introducing many new relation calls within the conjunction.
We believe this is the core problem of CPD which limits its power when applied to \mk.
The strategy of unfolding atoms from left to right is reasonable in the context of \pro because it mimics the way programs in \pro execute.
Special care should be taken when unfolding non-leftmost atoms in \pro: one should ensure that it does not duplicate code, as well as that no side-effects are done out of order~\cite{nonleftmost, leuschel2014fast}.
However in \mk leftmost unfolding often leads to larger global control trees and, as a result, bigger, less efficient programs.
On the contrary, according to the denotational semantics, the results of evaluation of a \mk program do not depend on the order of relation calls (atoms) within conjunctions, thus we believe a better result can be achieved by selecting a relation call which can restrict the number of branches in the tree.
We describe our approach, which implements this idea, in the next section.

\newcommand{\code}[1]{\texttt{#1}}

\section{Conservative Partial Deduction}
\label{conspd}

In this section, we describe a novel approach to relational program specialization.
This approach draws inspiration from both conjunctive partial deduction and supercompilation.
The aim was to create a specialization algorithm which would be simpler than conjunctive partial deduction and use properties of \mk to improve the performance of the input programs.

\begin{figure}[!t]
  \centering
  \includegraphics[width=\textwidth]{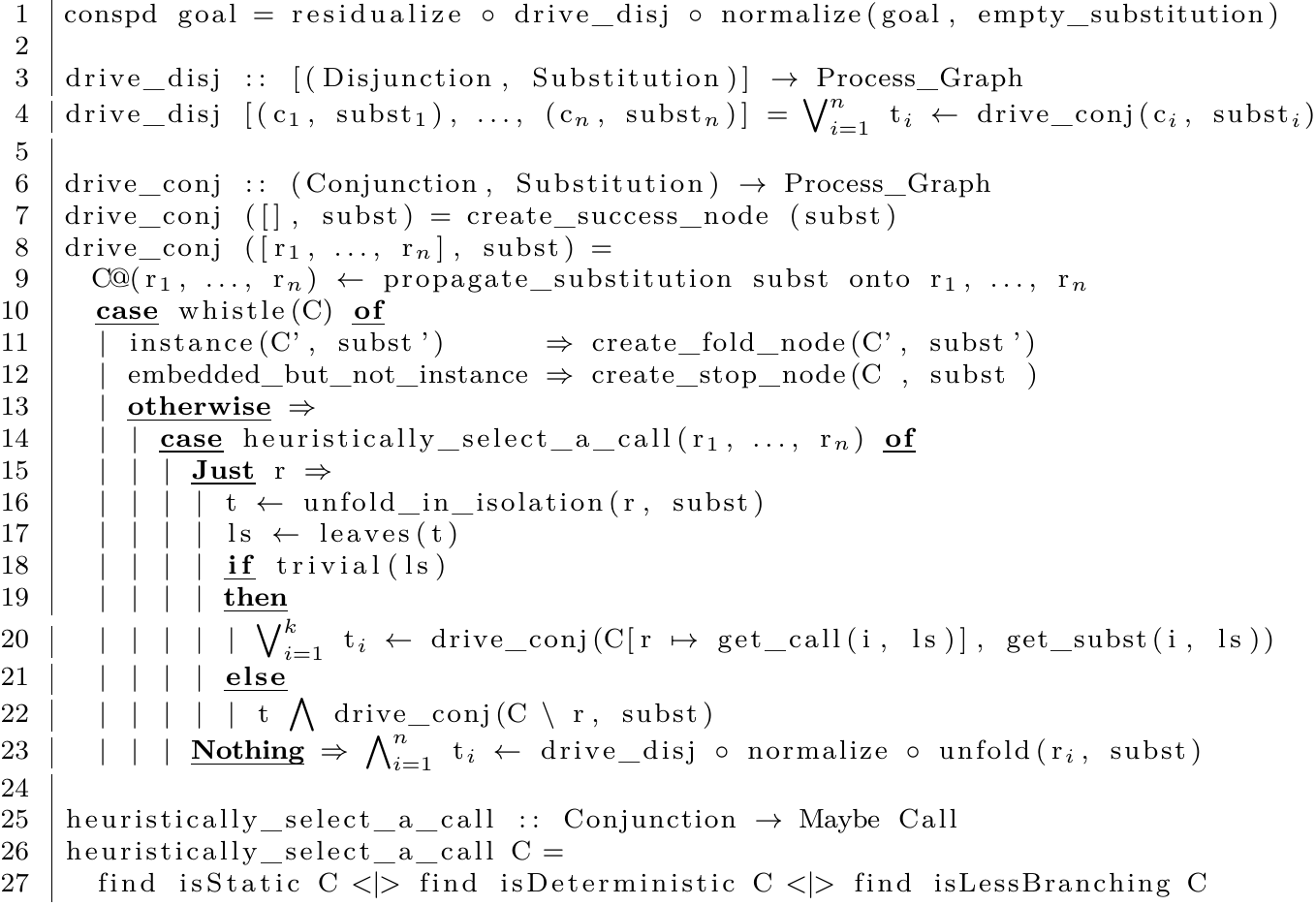}
  \caption{Conservative partial deduction pseudo code}
  \label{fig:ncpd-pseudo}
\end{figure}

The algorithm pseudocode is shown in Fig.~\ref{fig:ncpd-pseudo}.
We use the following notation in the pseudocode.
The circle, $\circ$, represents function composition.
The \emph{at} sign, @, is used to name a pattern matched value.
The arrow, $\leftarrow$, is used to bind a variable on the left-hand side.
The arrow can also be used to pattern match the value on the right-hand side (see line 9).
We use symbols $\bigwedge$ and $\bigvee$ to represent creating, respectively, a conjunction and a disjunction node in a process graph.
The data type \code{Maybe a = Just a | Nothing} is taken from \haskell and represents optional value.
The binary function \verb!x <|> y! combines two optional values: \verb!Just x <|> y = Just x!, while \verb!Nothing <|> y = y!.

Functions \code{drive\_disj} and \code{drive\_conj} describe how to process disjunctions and conjunctions respectively.
Hereafter, we consider all goals and relation bodies to be in \emph{canonical normal form}~---~a disjunction of conjunctions of either calls or unifications.
Moreover, we assume all fresh variables to be introduced into the scope and all unifications to be computed at each step.
Thus driving is declared to be the function \code{drive\_disj} (line 4).

A driving process (along with generalization and folding) creates a process graph, from which a residual program is later created.
The process graph is meant to mimic the execution of the input program.
The nodes of the process graph include a \emph{configuration} which describes the state of program evaluation at some point.
In our case a configuration is a conjunction of relation calls.
The substitution computed at each step is also stored in the graph node, although it is not included in the configuration.
This means that only the goal, and not the substitution, is passed into the \emph{whistle} to determine potential non-termination.

Residualization is done by the function \code{residualize}.
Residualization traverses the process graph and generates the \mk goal as well as new relations whenever needed.
A conjunction is created from a conjunction node, a disjunction --- from a disjunction node, while substitutions are generated into a conjunction of unifications.
Transient nodes --- the nodes which have a single child node --- correspond to intermediate functions in residual programs and are removed during residualization to improve the performance of programs.
We also employ \emph{redundant argument filtering} as described in~\cite{leuschel1996redundant}.

Those disjuncts in which unifications fail are removed while driving.
Each other disjunct takes the form of a possibly empty conjunction of relation calls accompanied with a substitution computed from unifications.
Any \mk term can be trivially transformed into the described form.
The function \code{normalize} in Fig.~\ref{fig:ncpd-pseudo} is assumed to perform term normalization.
The code is omitted for brevity but can be found in the implementation of the approach on Github\footnote{The project repository: \url{https://github.com/kajigor/uKanren_transformations/}. Access date: 28.02.2021}.


There are several core ideas behind this algorithm.
The first is to select an arbitrary relation to unfold, not necessarily the leftmost which is safe.
The second idea is to use a heuristic which decides if unfolding a relation call can lead to discovery of contradictions between conjuncts which in turn leads to restriction of the answer set at specialization-time (line 14; \code{heuristically\_select\_a\_call} stands for heuristics combination, see Section~\ref{sec:heurictic} for details).
If those contradictions are found, then they are exposed by considering the conjunction as a whole and replacing the selected relation call with the result of its unfolding thus \emph{joining} the conjunction back together instead of using \emph{split} as in CPD (lines 15--22).
Joining instead of splitting is why we call our transformer \emph{conservative} partial deduction.
Finally, if the heuristic fails to select a potentially good call, then the conjunction is split into individual calls which are driven in isolation and are never joined (line 23).

When the heuristic selects a call to unfold (line 15), a process tree is constructed for the selected call \emph{in isolation} (line 16).
This is done by driving the call until all leaves of the process tree are either substitutions, failures or recursive calls to the relations unfolded within the tree (example process tree is provided in Figure~\ref{fig:and}).
No folding is performed by \code{unfold\_in\_isolation} and each recursive call is unfolded at most once.
The leaves of the computed tree are examined.
If all leaves are either computed substitutions or are instances of some relations accompanied with non-empty substitutions, then the leaves are collected and each of them replaces the considered call in the root conjunction (line 20).
If the selected call does not suit the criteria, the results of its unfolding are not propagated onto other relation calls within the conjunction, instead, the next suitable call is selected (line 22).
According to the denotational semantics of \mk it is safe to compute individual conjuncts in any order, thus it is okay to drive any call and then propagate its results onto the other calls.


This process creates branchings whenever a disjunction is examined (lines 4, 20).
When a goal is fully computed to a substitution (line 7), then a success node is created and driving stops.
At each step, we make sure to not drive a conjunction which has already been examined.
To do this, we check if the current conjunction is a renaming of any other configuration in the graph (line 11).
If it is, then we fold by creating a special node which is then residualized into a call to the corresponding relation.

In this approach, we do not generalize in the same fashion as CPD or supercompilation.
This decision was motivated by keeping the complexity of the approach to the minimum.
Our conjunctions are always split into individual calls and are joined back together only if it is meaningful, for example, leads to contradictions.
If the need for generalization arises, i.e. homeomorphic embedding of conjunctions~\cite{de1999conjunctive} is detected, then we immediately stop driving this conjunction (line 12).
When residualizing such a conjunction, we just generate a conjunction of calls to the input program before specialization.



\subsection{Unfolding}

Unfolding in our case is done by substitution of some relation call by its body with simultaneous normalization and computation of unifications.
The unfolding itself is straightforward; however it is not always clear what to unfold and when to \emph{stop} unfolding.
Unfolding in the context of specialization of functional programming languages, as well as inlining in specialization of imperative languages, is usually considered to be safe from the residual program efficiency point of view.
It may only lead to code explosion or code duplication which is mostly left to a target program compiler optimization or even is out of consideration at all if a specializer is considered as a standalone tool~\cite{jonesbook}.

Unfortunately, this is not the case for the specialization of a relational programming language.
Unlike functional and imperative languages, in logic and relational programming languages unfolding may easily affect the target program's efficiency~\cite{leuschel2002logic, gallagher1993tutorial}.
Unfolding too much may create extra unifications, which is by itself a costly operation, or even introduce duplicated computations by propagating the results of unfolding onto neighbouring conjuncts.

There is a fine edge between too much unfolding and not enough unfolding.
The former is maybe even worse than the latter.
We believe that the following heuristic provides a reasonable approach to unfolding control.

\subsection{Less Branching Heuristic}
\label{sec:heurictic}

This heuristic is aimed at selecting a relation call within a conjunction which is both safe to unfold and may lead to discovering contradictions within the conjunction.
An unsafe unfolding leads to an uncontrollable increase of the number of relation calls in a conjunction.
It is best to first unfold those relation calls which can be fully computed up to substitutions.

We deem every static (non-recursive) conjunct to be safe because they never lead to growth in the number of conjunctions.
Those calls which unfold deterministically, meaning there is only one disjunct in the unfolded relation, are also considered to be safe.

Those relation calls which are neither static nor deterministic are examined with what we call the \emph{less-branching} heuristic.
It identifies the case when the unfolded relation contains fewer disjuncts than it could possibly have.
This means that we found some contradiction, some computations were gotten rid of, and thus the answer set was restricted, which is desirable when unfolding.
To compute this heuristic we precompute the maximum possible number of disjuncts in each relation and compare this number with the number of disjuncts when unfolding a concrete relation call.
The maximum number of disjuncts is computed by unfolding the body of the relation in which all relation calls were replaced by a unification which always succeeds.



The pseudocode describing our heuristic is shown in Fig.~\ref{fig:ncpd-pseudo} (lines 25-27).
Selecting a good relation call can fail.
We express this by using \verb!Maybe! data type for the result (line 25).
The implementation works as follows: we first select those relation calls which are static, and only if there are none, we proceed to consider deterministic unfoldings and then we search for those which are less branching.
We believe this heuristic provides a good balance in unfolding.




\section{Example}
\label{example}

In this section we demonstrate by example how conservative partial deduction works.
The example program is a relational interpreter of propositional formulas under given variable assignments.
The complete code of the example program is provided in Listing~\ref{eval:whole}.

The relation \lstinline{eval$^o$} has three arguments.
The first argument, \lstinline{s}, is a list of boolean values which plays the role of variable assignments.
The $i$-th value of the substitution is the value of the $i$-th variable.
The second argument, \lstinline{fm}, is a formula with the following abstract syntax.
A formula is either a \emph{variable} represented with a Peano number, a \emph{negation} of a formula, a \emph{conjunction} of two formulas or a \emph{disjunction} of two formulas.
The third argument, \lstinline{res}, is the value of the formula under the given assignment.

The relation \lstinline{eval$^o$} is in canonical normal form: it is a single disjunction which consists of 4 conjunctions of unifications and relation calls, and all its fresh variables are introduced at the top level.
The unification in each conjunction determines the shape of the input formula and binds variables to the corresponding subformulas.
For each of the subformulas, the relation \lstinline{eval$^o$} is called recursively.
Then the results of the evaluation of the subformulas are combined by the corresponding boolean connective to get the result for the input formula.
For example, when the formula is a conjunction of two subformulas \lstinline{x} and \lstinline{y}, the results of their execution \lstinline{v} and \lstinline{w} are combined by the call to the relation \lstinline{and$^o$} to compute the result: \lstinline{and$^o$ v w res}.
If the input formula is a variable, then its value is looked up in the substitution list by means of the relation \lstinline{elem$^o$}.

\begin{figure*}[!t]
  \centering
  \begin{minipage}{0.95\textwidth}
    \begin{lstlisting}[label={eval:whole}, caption={Evaluator of propositional formulas}, captionpos=b, frame=tb]
  let rec eval$^o$ s fm res = conde [fresh (x y z v w) (
      ( fm === conj x y /\ eval$^o$ s x v /\  eval$^o$ s y w /\  and$^o$ v w res );
      ( fm === disj x y /\ eval$^o$ s x v /\  eval$^o$ s y w /\  or$^o$   v w res );
      ( fm === neg x    /\ eval$^o$ s x v /\  not$^o$ v res );
      ( fm === var v    /\ elem$^o$ s v res ))]

  let not$^o$  x y = nand$^o$ x x y
  let or$^o$   x y z = nand$^o$ x x xx /\  nand$^o$ y y yy /\ nand$^o$ xx yy z
  let and$^o$ x y z = nand$^o$ x y xy /\   nand$^o$ xy xy z
  let nand$^o$ a b c = conde [
      ( a === false /\ b === false /\ c === true );
      ( a === false /\ b === true  /\ c === true );
      ( a === true  /\ b === false /\ c === true );
      ( a === true  /\ b === true  /\ c === false )]

  let elem$^o$ n s v = conde [ fresh (h t m) (
    ( n === zero /\ s === h : t /\ h === v );
    ( n === succ m /\ s === h : t /\ elem$^o$ m t v ))]


    \end{lstlisting}
  \end{minipage}
\end{figure*}

Consider the goal \lstinline{fresh (s fm) (eval$^o$ s fm true)}.
The partially constructed process graph for this goal is presented in Fig.~\ref{fig:evalTree}.
The following notation is used.
Rectangle nodes contain configurations, while diamond nodes correspond to splitting.
Each configuration contains a goal along with a substitution in angle brackets.
To visually differentiate constructors from variables within goals, we made them bold.
For brevity, we only put a fragment of the substitution computed at each step in the corresponding node.
The leaf node which contains only the substitution and no goal in its configuration is a success node.
The call selected to be unfolded is underlined in a conjunction.
Nodes corresponding to failures are not present within the process graph.
Dashed arrows mark renamings.

\begin{figure}[!t]
  \centering
  \begin{minipage}{0.95\textwidth}
    \begin{tikzpicture}[
  processTree]

  \node(0) {\underline{\rel{eval}{\texttt{fm s \textbf{true}}}}};

  \node(00)[below of=0, xshift=-7.5cm, yshift=0.6cm, anchor=north] {
    \rel{eval}{\texttt{x s a}} \conj \\
    \rel{eval}{\texttt{y s b}} \conj \\
    \underline{\rel{and}{\texttt{a b \textbf{true}}}} \\
    \subst{\texttt{fm} $\to$ \texttt{\textbf{conj} x y}}};

  \node(000)[below of=00, anchor=north] {
    \rel{eval}{\texttt{x s \textbf{true}}} \conj \\
    \rel{eval}{\texttt{y s \textbf{true}}} \\
    \subst{\texttt{a} $\to$ \texttt{\textbf{true},} \\ \texttt{b} $\to$ \texttt{\textbf{true}}}};

  \node(0000)[diamond,below of=000, yshift=0.5cm,anchor=north] { \conj };

  \node(00000)[below of=0000, xshift=-1.7cm, yshift=1cm, anchor=north] {
          \rel{eval}{\texttt{x s \textbf{true}}}};
  \node(00001)[below of=0000, xshift=1.7cm, yshift=1cm, anchor=north] {
          \rel{eval}{\texttt{y s \textbf{true}}}};

  \node(01)[below of=0, yshift=0.6cm, anchor=north] {
    \rel{eval}{\texttt{x s a}} \conj \\
    \rel{eval}{\texttt{y s b}} \conj \\
    \underline{\rel{or}{\texttt{a b \textbf{true}}}} \\
    \subst{\texttt{fm} $\to$ \texttt{\textbf{disj} x y}}};

  \node(010)[below of=01, xshift=-3.6cm, anchor=north] {
      \rel{eval}{\texttt{x s \textbf{true}}} \conj \\
      \rel{eval}{\texttt{y s \textbf{true}}} \\
      \subst{\texttt{a} $\to$ \texttt{\textbf{true},} \\ \texttt{b} $\to$ \texttt{\textbf{true}}}};

  \node(0100)[below of=010, anchor=north, draw=none, fill=none, yshift=0.5cm]{...};
  \node(011)[below of=01, anchor=north] {
      \rel{eval}{\texttt{x s \textbf{true}}} \conj \\
      \rel{eval}{\texttt{y s \textbf{false}}} \\
      \subst{\texttt{a} $\to$ \texttt{\textbf{true},} \\ \texttt{b} $\to$ \texttt{\textbf{false}}}};

  \node(0110)[diamond,below of=011, yshift=0.5cm,anchor=north] { \conj };

  \node(01100)[below of=0110, xshift=-1.7cm, yshift=1cm, anchor=north] {
          \rel{eval}{\texttt{x s \textbf{true}}}};
  \node(01101)[below of=0110, xshift=1.7cm, yshift=1cm, anchor=north] {
          \rel{eval}{\texttt{y s \textbf{false}}}};
  \node(011010)[below of=01101, xshift=-1.5cm, yshift=1.5cm, anchor=north, draw=none, fill=none] {...};
  \node(011011)[below of=01101, xshift=-0.5cm, yshift=1.5cm, anchor=north, draw=none, fill=none] {...};
  \node(011012)[below of=01101, xshift=0.5cm,  yshift=1.5cm, anchor=north, draw=none, fill=none] {...};
  \node(011013)[below of=01101, xshift=1.5cm,  yshift=1.5cm, anchor=north, draw=none, fill=none] {...};
  \node(012)[below of=01, xshift=3.7cm, anchor=north] {
      \rel{eval}{\texttt{x s \textbf{false}}} \conj \\
      \rel{eval}{\texttt{y s \textbf{true}}} \\
      \subst{\texttt{a} $\to$ \texttt{\textbf{false},} \\ \texttt{b} $\to$ \texttt{\textbf{true}}}};

  \node(0120)[below of=012, anchor=north, draw=none, fill=none, yshift=0.5cm]{...};

  \node(02)[below of=0, xshift=7.5cm, yshift=0.6cm, anchor=north] {
    \rel{eval}{\texttt{x s a} \conj \\
    \underline{\rel{not}{\texttt{a \textbf{true}}}} \\
    \subst{\texttt{fm} $\to$ \texttt{\textbf{neg} x}}}};

  \node(020)[below of=02,yshift=-0.23cm, anchor=north] {
    \rel{eval}{\texttt{x s \textbf{false}}} \\
    \subst{\texttt{a} $\to$ \texttt{\textbf{false}}}};

  \node(03)[below of=0, xshift=11.5cm, yshift=0.6cm, anchor=north] {
    \rel{elem}{\texttt{x s \textbf{true}}} \\
    \subst{\texttt{fm} $\to$ \texttt{\textbf{var} v}}
  };

  \node(030)[below of=03, xshift=-1cm, anchor=north, yshift=-0.5cm]{
    \subst{\texttt{s} $\to$ \texttt{h:t}, \\ \texttt{x} $\to$ \texttt{\textbf{zero}} \\ \texttt{h} $\to$ \texttt{\textbf{true}}}
  };
  \node(031)[below of=03, xshift=2cm, anchor=north, yshift=-0.5cm]{
    \rel{elem}{\texttt{y t \textbf{true}}} \\
    \subst{\texttt{s} $\to$ \texttt{h:t}, \\ \texttt{x} $\to$ \texttt{\textbf{succ} y}}};

  \draw [->] (0.south west) to (00.north east);
  \draw [->] (0) to (01);
  \draw [->] (0) to (02.north west);
  \draw [->] (0.south east) to (03.north west);
  \draw [->] (00) to (000);
  \draw [->] (000) to (0000);
  \draw [->] (0000) to (00000);
  \draw [->] (0000) to (00001);
  \draw [->] (01) to (010);
  \draw [->] (01) to (011);
  \draw [->] (01) to (012);
  \draw [->] (010) to (0100);
  \draw [->] (011) to (0110);
  \draw [->] (0110) to (01100);
  \draw [->] (0110) to (01101);
  \draw [->] (01101) to (011010);
  \draw [->] (01101) to (011011);
  \draw [->] (01101) to (011012);
  \draw [->] (01101) to (011013);
  \draw [->] (012) to (0120);
  \draw [->] (02) to (020);
  \draw [->] (03) to (030);
  \draw [->] (03) to (031);

  \draw [dashed,<-] ($(0.south west)+(0.1,0)$) .. controls +(-5,-2.5) and +(-4,0) .. (01100.west);
  \draw [dashed,->] (00000.north west) to [out=90,in=180] ($(0.west)$);
  \draw [dashed,->] (00001) to [out=90,in=-135] ($(0.west)+(0,-0.1)$);

  \draw[dashed,->] (020.south) to [out=-90,in=0] (01101.east);
  \draw[dashed,->] (031.north east) to [out=90,in=0] (03.south east);

\end{tikzpicture}
  \end{minipage}
  \caption{Partially constructed process graph for the relation eval$^o$.}
  \label{fig:evalTree}
\end{figure}
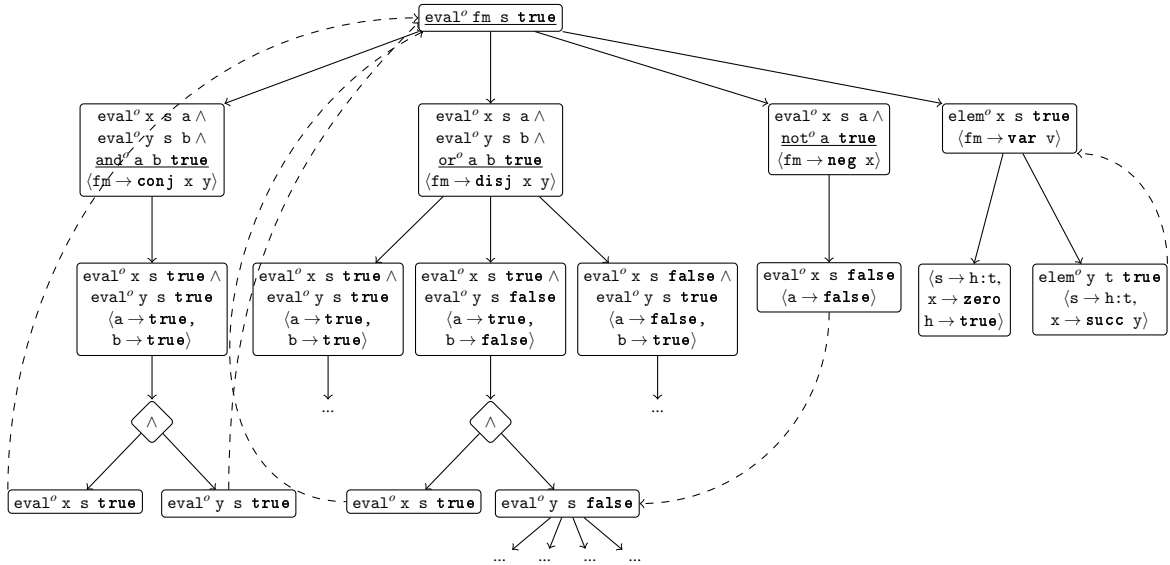

\begin{figure}[!t]
  \centering
  \begin{minipage}{0.95\textwidth}
    \begin{tikzpicture}[
  processTree,
  anchor=center,
  node distance=2.5cm]
  \node(0) {\rel{and}{\texttt{x y} \texttt{\textbf{true}}}};
  \node(00) [below of=0, yshift=1cm] {\rel{nand}{\texttt{x y xy} \conj \rel{nand}{\texttt{xy xy \texttt{\textbf{true}}}}}};
  \node(000) [below of=00, xshift=-10cm] {
    \rel{nand}{\texttt{\textbf{true true true}}} \\ \subst{\texttt{x} $\to$ \texttt{\textbf{false}}, \\ \texttt{y} $\to$ \texttt{\textbf{false}}, \\ xy $\to$ \texttt{\textbf{true}}}};
  \node(0000) [below of=000, draw=none] {fail};
  \node(001) [below of=00, xshift=-3cm] {
    \rel{nand}{\texttt{\textbf{true true true}}} \\ \subst{\texttt{x} $\to$ \texttt{\textbf{false}}, \\ \texttt{y} $\to$ \texttt{\textbf{true}}, \\ xy $\to$ \texttt{\textbf{true}}}};
  \node(0010) [below of=001, draw=none] {fail};

  \node(002) [below of=00, xshift=3cm] {
    \rel{nand}{\texttt{\textbf{true true true}}} \\ \subst{\texttt{x} $\to$ \texttt{\textbf{true}}, \\ \texttt{y} $\to$ \texttt{\textbf{false}}, \\ xy $\to$ \texttt{\textbf{true}}}};
  \node(0020) [below of=002, draw=none] {fail};

  \node(003) [below of=00, xshift=10cm] {
    \rel{nand}{\texttt{\textbf{false false true}}} \\ \subst{\texttt{x} $\to$ \texttt{\textbf{true}}, \\ \texttt{y} $\to$ \texttt{\textbf{true}}, \\ xy $\to$ \texttt{\textbf{false}}}};
  \node(0030) [below of=003] { \subst{\texttt{x} $\to$ \texttt{\textbf{true}}, \texttt{y} $\to$ \texttt{\textbf{true}}, xy $\to$ \texttt{\textbf{false}}}};

  \draw[->] (0) to (00);

  \draw[->] (00) to (000);
  \draw[->] (00) to (001);
  \draw[->] (00) to (002);
  \draw[->] (00) to (003);

  \draw[->] (000) to (0000);
  \draw[->] (001) to (0010);
  \draw[->] (002) to (0020);
  \draw[->] (003) to (0030);

\end{tikzpicture}
  \end{minipage}
  \caption{Unfolding of and$^o$.}
  \label{fig:and}
\end{figure}
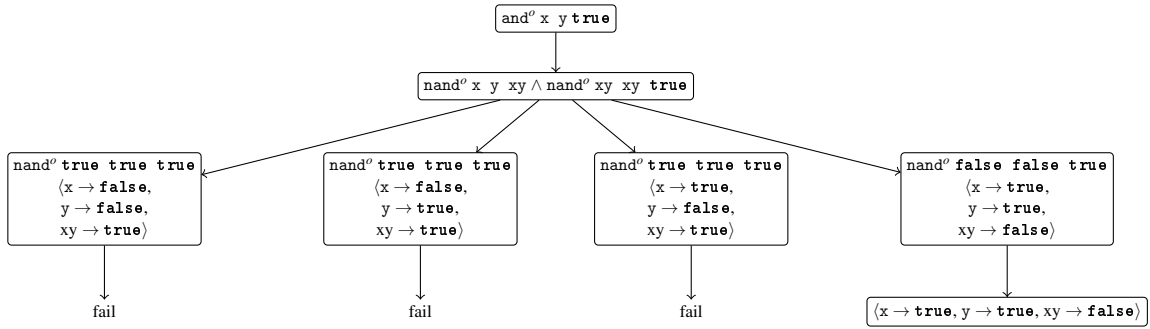

Conservative partial deduction starts by unfolding the single relation call in this goal once.
Besides introducing fresh variables into the context, unifications in each conjunct are computed into substitutions.
This produces 4 branches, each of which is processed further.

Consider the first branch, in which the input formula \lstinline{fm} is a conjunction of subformulas \lstinline{x} and \lstinline{y}.
First, each relation call within a conjunction is examined separately to select one of the calls to unfold.
To do so, we unfold each of them in isolation and use the less branching heuristic.
Both recursive calls to \lstinline{eval$^o$} are done with three distinct fresh variables, and are not selected according to the less branching heuristic.
By unfolding the call to \lstinline{and$^o$} several times, we determine that it has a single non-failing branch (see Fig.~\ref{fig:and}), which is less than the same relation would have if called on all free and distinct variables, thus this call is selected to be unfolded.
The result of unfolding the call is a single substitution which associates the variables \lstinline{a} and \lstinline{b} with \lstinline{true}.
By applying the computed substitution to the goal we get a conjunction of two calls to the \lstinline{eval$^o$} relation with the last argument being \lstinline{true}.
This conjunction embeds the root goal, thus we split the conjunction and both calls become leaves which rename the root.
This finishes the processing of the first branch.

Consider the second branch, in which the input formula \lstinline{fm} is a disjunction of subformulas \lstinline{x} and \lstinline{y}.
Similarly to the first branch, the heuristic selects the call to the boolean relation to be unfolded which produces three possible substitutions for variables \lstinline{a} and \lstinline{b}.
The substitutions are propagated into the goals and three branches, each of which embeds the root goal, are added into the process graph.
Each goal is then split, and the calls with the last argument being \lstinline{true} rename the root.
Finally, the call \lstinline{eval$^o$ y s false} is processed, which creates 4 branches similar to the branches of the root goal.

\begin{figure*}[!t]
  \centering
  \begin{minipage}{0.65\textwidth}
    \begin{lstlisting}[label={eval:conspd}, caption={Specialized evaluator of propositional formulas}, captionpos=b, frame=tb]
  let rec eval$^o_{true}$ s fm = conde [fresh (x y z v w) (
      ( fm === conj x y /\ eval$^o_{true}$ s x /\  eval$^o_{true}$ s y );
      ( fm === disj x y /\ (conde [
          ( eval$^o_{true}$ s x /\    eval$^o_{true}$ s y );
          ( eval$^o_{true}$ s x /\    eval$^o_{false}$ s y );
          ( eval$^o_{false}$ s x /\    eval$^o_{true}$ s y );
      ]);
      ( fm === neg x /\ eval$^o_{false}$ s x v );
      ( fm === var v /\ elem$^o_{true}$ s v ))]

  let rec eval$^o_{false}$ s fm = conde [fresh (x y z v w) (
      ( fm === conj x y /\ (conde [
          ( eval$^o_{false}$ s x /\    eval$^o_{false}$ s y );
          ( eval$^o_{true}$ s x /\    eval$^o_{false}$ s y );
          ( eval$^o_{false}$ s x /\    eval$^o_{true}$ s y );
      ]);
      ( fm === disj x y /\ eval$^o_{true}$ s x /\  eval$^o_{true}$ s y );
      ( fm === neg x /\ eval$^o_{true}$ s x v );
      ( fm === var v /\ elem$^o_{false}$ s v ))]

  let elem$^o_{true}$ n s = conde [ fresh (h t m) (
      ( n === zero /\ s === true : t );
      ( n === succ m /\ s === h : t /\ elem$^o_{true}$ m t ))]

  let elem$^o_{false}$ n s = conde [ fresh (h t m) (
      ( n === zero /\ s === false : t );
      ( n === succ m /\ s === h : t /\ elem$^o_{false}$ m t ))]


    \end{lstlisting}
  \end{minipage}
\end{figure*}

The third branch is driven until a call with the last argument being \lstinline{false} is encountered.
Since it renames one of the nodes which is already present in the process graph, we stop exploring this branch and add the back edge.

The last branch, in which the input formula \lstinline{fm} is a variable \lstinline{v}, contains the single call to the relation \lstinline{elem$^o$}.
The unfolding of this call produces two leaves: a success node and a renaming of the parent node.
This finishes the construction of the process graph.

The process graph is then residualized into a specialized version of the \lstinline{eval$^o$} relation (see Listing~\ref{eval:conspd}).
This program does not contain any calls to boolean connectives.
Neither does the program contain the original, not specialized, relation \lstinline{eval$^o$}.

It is worth noting that the result produced by the Conservative Partial Deduction is not ideal.
For example, in the definition of the \lstinline{eval$^o_{true}$}, when the input formula \lstinline{fm} is a disjunction of subformulas \lstinline{x} and \lstinline{y}, the recursive call \lstinline{eval$^o_{true}$ s x } is done twice in two disjuncts.
The ideal version of the relation \lstinline{eval$^o_{true}$} should contain this recursive call only once.
This can be done, for example, by common subexpression elimination~\cite{muchnick1997advanced}.
However, ideally, \lstinline{y} should not be evaluted at all, since the value of the formula \lstinline{fm} does not depend on it.
It is unclear if and how this kind of transformation can be done automatically.
Such transformation would require, first, realising that a disjunction of two relation calls \lstinline{eval$^o_{true}$ s y} and \lstinline{eval$^o_{false}$ s y} exhaust all possible values for \lstinline{y}.
Secondly, the transformation would have to examine if a relation restricts values of a given argument regardless of the other arguments' values.

\section{Evaluation}
\label{evaluation}

We implemented\footnote{The project repository: \url{https://github.com/kajigor/uKanren_transformations/}. Access date: 28.02.2021} the conservative partial deduction for \mk and compared it with the \ecce partial deduction system.
\ecce is designed for the \pro programming language and cannot be directly applied for programs written in \mk.
Nevertheless, the languages show resemblance, and it is valuable to check if the existing methods for \pro can be used directly in the context of relational programming.
To be able to compare our approach with \ecce, we converted each input program first to the pure subset of \pro, then specialized it with \ecce, and then we converted the result back to \mk.
The conversion to \pro is a simple syntactic conversion. In the conversion from \pro to \mk, for each Horn clause a conjunction is generated in which unifications are placed before any relation call.
All programs are run as \mk programs in our experiments.
In the final subsection we discuss some limitations of both our approach and ECCE.

We chose two problems for our study: evaluation of a subset of propositional formulas and typechecking for a simple language.
The problems illustrate the approach of using relational interpreters to solve search problems~\cite{lozov2019relational}.
For both these problems we considered several possible implementations in \mk which highlight different aspects relevant in specialization.

The \lstinline{eval$^o$} relation implements an evaluator of a subset of propositional formulas described in Section~\ref{example}.
We consider four different implementations of this relation to explore how the way the program is implemented can affect the quality of specialization.
Depending on the implementation, \ecce generates programs of varying performance, while the execution times of the programs generated by our approach are similar.

The \lstinline{typecheck$^o$} relation implements a typechecker for a tiny expression language.
We consider two different implementations of this relation: one written by hand and the other generated from a functional program, which implements the typechecker, as described in~\cite{lozov2019relational}.
We demonstrate how much these implementations differ in terms of performance before and after specialization.

In this study we measured the execution time for the sample queries, averaging them over multiple runs.
All examples of \mk relations in this paper are written in \oc.
The queries were run on a laptop running Ubuntu 18.04 with quad core Intel Core i5 2.30GHz CPU and 8 GB of RAM.

The tables and graphs use the following denotations.
\emph{Original} represents the execution time of a program before any transformations were applied; \emph{ECCE}~--- of the program specialized by \ecce with the default conjunctive control setting; \emph{ConsPD}~--- of the program specialized by our approach.

\subsection{Evaluator of Logic Formulas}

The relation \lstinline{eval$^o$} describes an evaluation of a propositional formula under given variable assignments presented in Section~\ref{example}.
We specialize the \lstinline{eval$^o$} relation to synthesize formulas which evaluate to \lstinline{true}.
To do so, we run the specializer for the goal with the last argument fixed to \lstinline{true}, while the first two arguments remain free variables.
Depending on the way \lstinline{eval$^o$} is implemented, different specializers generate significantly different residual programs.

\subsubsection{The Order of Relation Calls}

One possible implementation of the \lstinline{eval$^o$} relation is presented in Listing~\ref{eval:last}.
Here the relation \lstinline{elem$^o$ s v res} unifies \lstinline{res} with the value of the variable \lstinline{v} in the list \lstinline{s}.
The relations \lstinline{and$^o$}, \lstinline{or$^o$}, and \lstinline{not$^o$} encode corresponding boolean connectives.

\begin{figure*}[!t]
  \centering
  \begin{minipage}{0.95\textwidth}
    \begin{lstlisting}[label={eval:last}, caption={Evaluator of formulas with boolean operation last}, captionpos=b, frame=tb]
  let rec eval$^o$ s fm res = conde [fresh (x y z v w) (
      (fm === conj x y /\ eval$^o$ s x v /\  eval$^o$ s y w /\  and$^o$ v w res);
      (fm === disj x y /\ eval$^o$ s x v /\  eval$^o$ s y w /\  or$^o$   v w res);
      (fm === neg x    /\ eval$^o$ s x v /\  not$^o$ v res));
      (fm === var v    /\ elem$^o$ s v res)]
    \end{lstlisting}
  \end{minipage}
  \begin{minipage}{0.95\textwidth}
    \begin{lstlisting}[label={eval:fst}, caption={Evaluator of formulas with boolean operation second}, captionpos=b, frame=tb]
  let rec eval$^o$ s fm res = conde [fresh (x y z v w) (
      (fm === conj x y /\ and$^o$ v w res /\  eval$^o$ s x v /\  eval$^o$ s y w);
      (fm === disj x y /\ or$^o$   v w res /\ eval$^o$ s x v /\  eval$^o$ s y w);
      (fm === neg x    /\ not$^o$ v res   /\ eval$^o$ s x v);
      (fm === var v    /\ elem$^o$ s v res))]
    \end{lstlisting}
  \end{minipage}
\end{figure*}

Note, the calls to boolean relations \lstinline{and$^o$}, \lstinline{or$^o$}, and \lstinline{not$^o$} are placed last within each conjunction.
This poses a challenge for the CPD-based specializers such as \ecce.
Conjunctive partial deduction unfolds relation calls from left to right, so when specializing this relation for running backwards (i.e. considering the goal \lstinline{eval$^o$ s fm true}), it fails to propagate the direction data onto recursive calls of \lstinline{eval$^o$}.
Knowing that \lstinline{res} is \lstinline{true}, we can conclude that the variables \lstinline{v} and \lstinline{w} have to be \lstinline{true} as well in the call \lstinline{and$^o$ v w res}.
There are three possible options for these variables in the call \lstinline{or$^o$ v w res} and one for the call \lstinline{not$^o$ v res}.
These variables are used in recursive calls of \lstinline{eval$^o$} and thus restrict the result of its execution.
CPD fails to recognize this, and thus unfolds recursive calls of \lstinline{eval$^o$} applied to fresh variables.
It leads to over-unfolding, large residual programs and poor performance.

The conservative partial deduction first unfolds those calls which are selected according to the heuristic.
Since exploring the implementations of boolean connectives makes more sense, they are unfolded before the recursive calls of \lstinline{eval$^o$}.
The way conservative partial deduction treats this program is the same as it treats the other implementation in which boolean connectives
are moved to the left, as shown in Listing~\ref{eval:fst}.
This program is easier for \ecce to specialize which demonstrates how unequal the behaviour of CPD for similar programs is.

\subsubsection{Unfolding of Complex Relations}

Depending on the way a relation is implemented, it may take a different number of driving steps to reach the point when any useful information is derived through its unfolding.
Partial deduction tries to unfold every relation call unless it is unsafe, but not all relation calls serve to restrict the search space and thus should be unfolded.
In the implementation of \lstinline{eval$^o$} boolean connectives can effectively restrict variables within the conjunctions and should be unfolded until they do.
But depending on the way they are implemented, the different number of driving steps should be performed for that.
The simplest way to implement these relations is by mimicking a truth table as demonstrated by the implementation of \lstinline{not$^o$} in Listing~\ref{not:table}.
It is enough to unfold such relation calls once to derive useful information about variables.

\begin{figure*}[!t]
  \centering
  \begin{minipage}{0.5\textwidth}
    \begin{lstlisting}[label={not:table}, caption={Implementation of boolean not$^o$ as a table}, captionpos=b, frame=tb]
      let not$^o$ x y = conde [
         (x === true /\ y === false;
          x === false /\ y === true)]
    \end{lstlisting}
  \end{minipage}
  \begin{minipage}{0.8\textwidth}
    \begin{lstlisting}[label={not:nando}, caption={Implementation of boolean operations via nand$^o$}, captionpos=b, frame=tb]
  let not$^o$   x y = nand$^o$ x x y
  let or$^o$   x y z = nand$^o$ x x xx /\  nand$^o$ y y yy /\ nand$^o$ xx yy z
  let and$^o$ x y z = nand$^o$ x y xy /\   nand$^o$ xy xy z
  let nand$^o$ a b c = conde [
    ( a === false /\ b === false /\ c === true );
    ( a === false /\ b === true  /\ c === true );
    ( a === true  /\ b === false /\ c === true );
    ( a === true  /\ b === true  /\ c === false)]
    \end{lstlisting}
  \end{minipage}
\end{figure*}

The other way to implement boolean connectives is to express them using a single basic boolean relation such as \lstinline{nand$^o$} which, in turn, has a table-based
implementation (see Listing~\ref{not:nando}). It takes several sequential unfoldings to derive that the variables \lstinline{v} and \lstinline{w} should
be \lstinline{true} when considering a call \lstinline{and$^o$ v w true} implemented via a basic relation.
Conservative partial deduction drives the selected call until it derives useful substitutions for the variables involved while CPD with deterministic unfolding may fail to do so.

\subsubsection{Evaluation Results}

\begin{figure}[!t]
  \centering
  \begin{subfigure}[c]{0.35\textwidth}
    \centering
    \begin{tabular}{e{1cm}||c|c|c}
               & Original & \ecce & ConsPD \\ \hline\hline
      \emph{FirstPlain} & 1.59s & 1.61s & 0.92s \\ \hline
      \emph{FirstNando} & 1.43s & 2.24s & 0.96s \\ \hline
      \emph{LastPlain}  & 0.98s & 1.43s & 0.97s \\ \hline
      \emph{LastNando}  & 1.09s & 1.54s & 0.91s
    \end{tabular}
  \end{subfigure}
  \hfill
  \begin{subfigure}[c]{0.58\textwidth}
    \includegraphics[width=\textwidth]{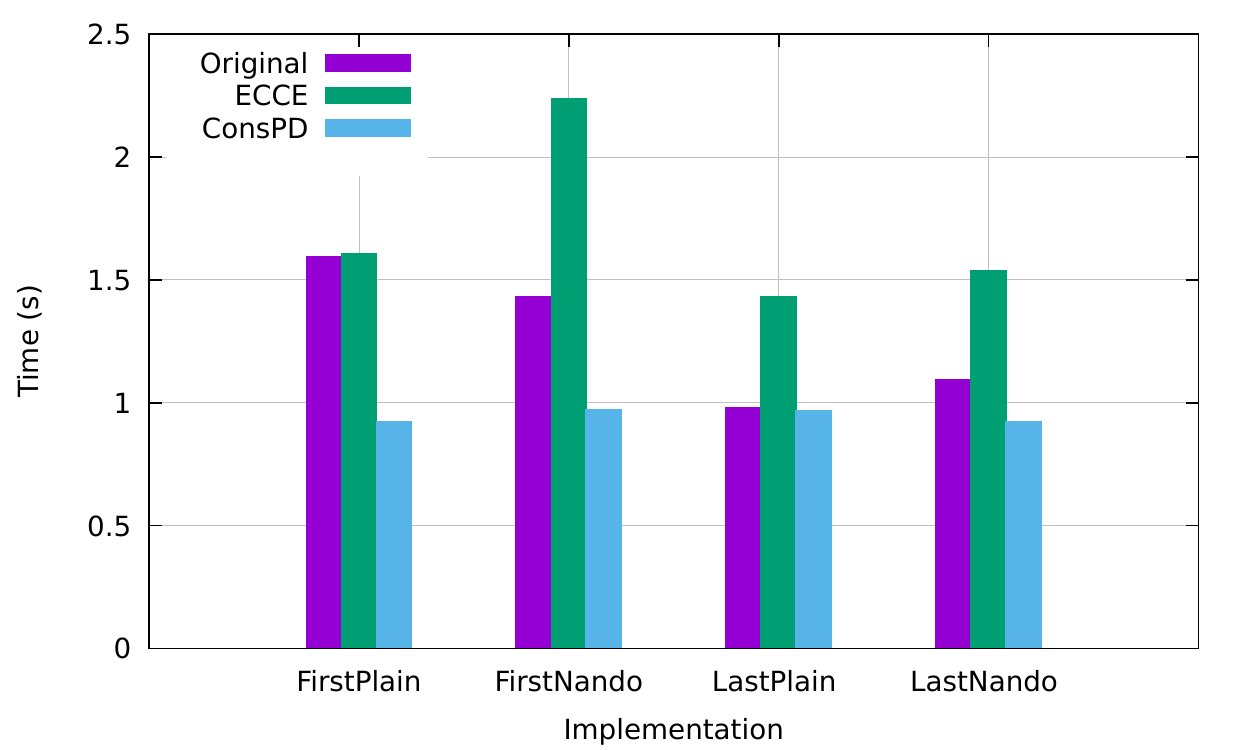}
  \end{subfigure}
  \caption{Execution time of eval$^o$}
  \label{fig:eval}
\end{figure}

In our study we considered 4 implementations of \lstinline{eval$^o$} summarised in the Table~\ref{tbl:eval}. They differ in the way the boolean connectives are implemented (see column \emph{Implementation}) and whether they are placed before or after the recursive calls to \lstinline{eval$^o$} (see column \emph{Placement}).
These four implementations are very different from the  standpoint of \ecce.
We measured the time necessary to generate $1000$ formulas over two variables which evaluate to \lstinline{true} (averaged over 10 runs).
The results are presented in Fig.~\ref{fig:eval}.

\begin{table}[!h]
    \centering
    \begin{tabular}{c||c||c}
                      & Implementation & Placement \\ \hline\hline
    \emph{FirstPlain} & table-based    & before \\ \hline
    \emph{LastPlain}  & table-based    & after  \\ \hline
    \emph{FirstNando} & via nand$^o$   & before \\ \hline
    \emph{LastNando}  & via nand$^o$   & after  \\
    \end{tabular}

  \caption{Different implementations of eval$^o$}
  \label{tbl:eval}
\end{table}

Conservative partial deduction generates  programs with comparable performance for all four implementations, while the quality of \ecce specialization differs significantly.
\ecce worsens performance for every implementation as compared to the original program.
ConsPD does not worsen performance for any implementation.
Its effect is most significant for the implementations in which the boolean connectives are placed first within conjunctions.

\subsubsection{The Order of Answers}

It is important to note that different implementations of the same \mk relation produce answers in different orders.
Nevertheless, since \mk search is complete, all answers will be found eventually.
Unfortunately, it is not guaranteed that the first 1000 formulas generated with different implementations of \lstinline{eval$^o$} will be the same.
For example, $983$ formulas are the same among the first $1000$ formulas generated by the Original \emph{FirstPlain} relation and the same relation after the ConsPD transformation.
At the same time, only $405$ formulas are the same between the Original and \ecce \emph{LastNando} relations.

The reason why implementations differ so much in the order of the answers stems from the canonical search strategy employed in \mk.
Most \mk implementations employ \emph{interleaving} search~\cite{10.1145/1090189.1086390} which is left-biased.
It means that the leftmost disjunct in a relation is being executed longer than the disjunct on the right.
This property is not local which makes it very hard to estimate the performance of a given relation.

In practice it means that if a specializer reorders disjuncts, then the performance of relations after specialization may be unpredictable.
For example, by putting the disjuncts of the \lstinline{eval$^o$} relation in the opposite order, one produces a relation which runs much faster than the original, but it generates completely different formulas at the same time.
Most of the first 1000 formulas in this case are multiple negations of a variable, while the original relation produces more diverse set of answers.
Computing a negation of a formula only takes one recursive \lstinline{eval$^o$} call thus finding such answers is faster than conjunctions and disjunctions.
Meanwhile, the formulas generated by the reordered relation are less diverse and may be of less interest.

Although neither \ecce nor ConsPD reorder disjuncts, they remove disjuncts which cannot succeed.
Thus they influence the order of answers and performance of relations.
Both methods reduce the number of unifications needed to compute each individual answer thus performing specialization.
In general, it is not possible to guarantee the same order of answers after specialization.
Exploring how different specializations influence the execution order is a fascinating direction for future research.

\subsection{Typechecker-Term Generator}

This relation implements a typechecker for a tiny expression language.
Being executed in the backward direction it serves as a generator of terms of the given type.
The abstract syntax of the language is presented below.
The variables are represented with de Bruijn indices, thus let-binding does not specify which variable is being bound.

\[\begin{array}{lllll}
  type \ term = &\ BConst \ of \ Bool &| \ IConst \ of \ Int &| \ Var \ of \ Int & \\
  & | \ term + term &| \ term * term &| \ term = term &| \ term < term \\
  &| \ \underline{let} \ term \ \underline{in} \ term
  &\multicolumn{2}{l}{| \ \underline{if} \ term \ \underline{then} \ term \ \underline{else} \ term} &
\end{array}\]

The typing rules are straightforward and are presented in Fig.~\ref{fig:typing}.
Boolean and integer constants have the corresponding types regardless of the environment.
Only terms of type integer can be added, multiplied or compared by the less-than operator.
Any terms of the same type can be checked for equality.
Addition and multiplication of two terms of suitable types have integer type, while comparisons have boolean type.
The if-then-else expression typechecks only if its condition is of type boolean, while both then- and else-branches have the same type.
An environment $\Gamma$ is a list, in which the $i$-th element is the type of the variable with the $i$-th de Bruijn index.
To typecheck a let-binding, first, the term being bound is typechecked and is added in the beginning of the environment $\Gamma$, and then the body is typechecked in the context of the new environment.
Typechecking a variable with the index $i$ boils down to getting an $i$-th element of the list.

\begin{figure}[!h]
  \setlength{\tabcolsep}{0.4cm}
  \centering
  \begin{tabular}{c c c}
    \infer[]{\Gamma \vdash IConst \ i : Int}{} &
    \infer[]{\Gamma \vdash BConst \ b : Bool}{}  &
    \infer[\Gamma \lbrack v \rbrack \equiv \tau]{\Gamma \vdash Var \ v : \tau}{} \vspace{0.5cm}
    \\
    \infer[]{\Gamma \vdash t + s : Int}{\Gamma \vdash t : Int, \Gamma \vdash  s : Int}  \vspace{0.5cm} &
    \infer[]{\Gamma \vdash t = s : Bool}{\Gamma \vdash t : \tau, \Gamma \vdash  s : \tau} &
    \infer[]{\Gamma \vdash \underline{let} \ v \ b : \tau}{\Gamma \vdash v : \tau_v, \ (\tau_v :: \Gamma) \vdash b : \tau}
      \\

      \infer[]{\Gamma \vdash t * s : Int}{\Gamma \vdash t : Int, \Gamma \vdash  s : Int}  &
    \infer[]{\Gamma \vdash t < s : Bool}{\Gamma \vdash t : Int, \Gamma \vdash  s : Int} \vspace{0.5cm} &
      \infer[]{\Gamma \vdash \underline{if} \ c \ \underline{then} \ t \ \underline{else} \ s : \tau}{\Gamma \vdash c : Bool, \Gamma \vdash t : \tau, \Gamma \vdash s : \tau}
  \end{tabular}
  \vspace{-0.3cm}
  \caption{Typing rules implemented in typecheck$^o$ relation}
  \label{fig:typing}
\end{figure}

\begin{figure}[!h]
  \begin{subfigure}[c]{0.55\textwidth}
    \centering
    \begin{tabular}{c||c|c|c}
                          & Original & \ecce & ConsPD  \\ \hline\hline
      \emph{Hand-written} & 0.92s    & 0.22s & 0.34s   \\ \hline
      \emph{Generated}    & 11.46s   & 0.38s & 0.29s
      \end{tabular}
  \end{subfigure}
  \hfill
  \begin{subfigure}[c]{0.45\textwidth}
    \includegraphics[width=\textwidth]{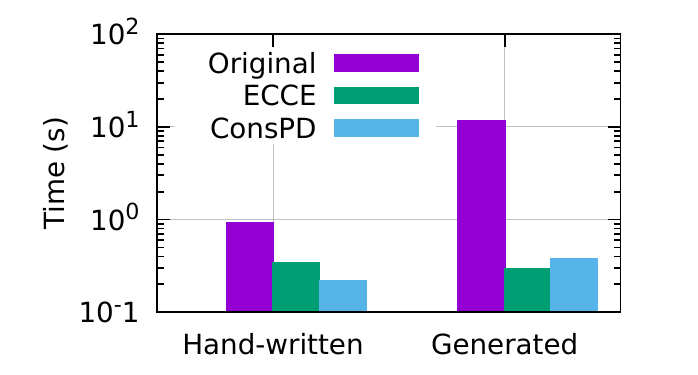}
  \end{subfigure}
  \caption{Execution  time of generating 1000 closed terms of type integer}
  \label{tbl:type}
\end{figure}

We compared two implementations of these typing rules.
The first one is obtained by unnesting of a functional program, which implements the typechecker, as described in~\cite{lozov2019relational} (\emph{Generated}).
It is worth noting that the unnesting introduces a lot of redundancy in the form of extra unifications and thus creates programs which are very inefficient.
Thus we contrast this implementation with the program hand-written in \oc (\emph{Hand-written}).
Each implementation has been specialized with ConsPD and \ecce.
We measured the time needed to generate 1000 closed terms of type integer (see Fig.~\ref{tbl:type}).

As expected, the generated program is far slower than the hand-written one.
The principal difference between these two implementations is that the generated program contains a certain redundancy introduced by unnesting.
For example, typechecking of the sum of two terms in the hand-written implementation consists of a single conjunction (see Listing~\ref{type:hand}) while the generated program is far more complicated and also uses a special relation \lstinline{typeEq$^o$} to compare types (see Listing~\ref{type:gen}).
This relation has to be unfolded early to determine types of the subterms, much like the boolean relations implemented via a basic relation had to be unfolded in the previous problem.

\begin{figure*}[!t]
  \centering
    \begin{lstlisting}[label={type:hand}, caption={A fragment of the hand-written typechecker}, captionpos=b, frame=tb]
  let rec typecheck$^o$ gamma term res = conde [
    ...
    fresh (x y) ((term === x + y /\
       typecheck$^o$ gamma x (some integer) /\
       typecheck$^o$ gamma y (some integer) /\
       res === (some integer)));
    ...]
    \end{lstlisting}
\end{figure*}

\begin{figure*}[!t]
  \centering
    \begin{lstlisting}[label={type:gen}, caption={A fragment of the generated typechecker}, captionpos=b, frame=tb]
let rec typecheck$^o$ gamma term res = conde [
  ...
  fresh (x y t1 t2) ((term === x + y /\
    conde [
      typecheck$^o$ gamma x none       /\ res === none;
      typecheck$^o$ gamma x (some t1) /\
        typecheck$^o$ gamma y none     /\ res === none;
      typecheck$^o$ gamma x (some t1) /\  typecheck$^o$ gamma y (some t2) /\
        typeEq$^o$ t1 integer true     /\ typeEq$^o$ t2 integer true /\
        res === (some integer);
    ])
  ...]
    \end{lstlisting}
\end{figure*}

Most redundancy of the generated program is removed by specialization with respect to the known type.
This is why both implementations have comparable speed after specialization.
\ecce shows bigger speedup for the hand-written program than ConsPD and vice versa for the generated implementation.
We believe that this difference can be explained by too much unfolding.
\ecce performs a lot of excessive unfolding for the generated program and only barely changes the hand-written program.
At the same time ConsPD specializes both implementations to comparable programs performing an average amount of unfolding.
This shows that the heuristic we presented gives more stable, although not the best, results.

\subsection{Discussion: Tupling and Deforestation}
\label{discussion}

\begin{figure*}[!t]
  \centering
  \begin{minipage}{0.7\textwidth}
    \begin{lstlisting}[label={doubleApp}, caption={Inefficient implementation of concatenation of three lists}, captionpos=b, frame=tb]
    let doubleAppend$^o$ x y z res = fresh (t) (
        append$^o$ x y t /\ append$^o$ t z res )

    let append$^o$ x y res = conde [ fresh (h t r) (
        ( x === [] /\ res === y );
        ( x === h : t /\ append$^o$ t y r /\ res === h : r ))]
    \end{lstlisting}
  \end{minipage}
\end{figure*}

\begin{figure*}[!t]
  \centering
  \begin{minipage}{0.7\textwidth}
    \begin{lstlisting}[label={maxlen}, caption={Inefficient implementation of maxLength$^o$}, captionpos=b, frame=tb]
    let maxLength$^o$ x m l = fresh (t) (
        max$^o$ x m /\ length$^o$ x l )

    let length$^o$ x l = conde [ fresh (h t r) (
        ( x === [] /\ l === zero );
        ( x === h : t /\ length$^o$ t r /\ l === succ r ))]

    let max$^o$ x m = max$_1^o$ x zero m
    let rec max$_1^o$ x n m = fresh (h t) ( conde [
        (x === [] /\ m === n);
        (x === h : t) /\ (conde [
          (le$^o$ h n true /\  max$_1^o$ t n m);
          (gt$^o$ h n true /\  max$_1^o$ t h m)])])
    \end{lstlisting}
  \end{minipage}
\end{figure*}

Tupling~\cite{pettorossi1984powerful, chin1993towards} and deforestation are among the important transformations conjunctive partial deduction is capable of.
Deforestation is often demonstrated by the \lstinline{doubleAppend$^o$} program which concatenates three lists by calling the concatenation relation \lstinline{append$^o$} twice in a conjunction (see Listing~\ref{doubleApp}).
The two calls to \lstinline{append} lead to double traversal of the first list, which is inefficient.
The program may be transformed in such a way so as to only traverse the first list once (see~\cite{de1999conjunctive} for details), which conjunctive partial deduction does.

Conjunctive partial deduction achieves this effect by considering the conjunction of two \lstinline{append$^o$} calls as a whole.
At the local control level, it first unfolds the leftmost call, propagates the computed substitutions onto the rightmost call, and then unfolds the rightmost call in the context of the substitutions.
When the first list is not empty, this leads to discovering a renaming of a conjunction of two \lstinline{append$^o$} calls.
By renaming this conjunction into a new predicate, deforestation is achieved in this example.

Unfortunately, conservative partial deduction does not succeed at this transformation on this example.
This happens because ConsPD splits the conjunction of two calls to \lstinline{append$^o$}, since none of them is selected by the less branching heuristic.
Splitting the conjunction leads to information loss and makes it so there is no renaming of the whole conjunction in the process graph.

A similar thing happens when considering the common example on which tupling is demonstrated in literature on CPD: the \lstinline{maxLength$^o$} program.
The original implementation of this program computes the maximum element of the list along with the length of the list by conjunction of two calls to the relations \lstinline{max$^o$} and \lstinline{length$^o$} respectively (see Listing.~\ref{maxlen}).
This implementation also traverses the input list twice when run in the forward direction.
By tupling, this program may be transformed so that the list is traversed once while both the maximum value and the length of the list are computed simultaneously, and CPD is capable to achieve this transformation with the default settings.
Conservative partial deduction also splits much too early and thus fails to yield any useful transformation for this program.

It is worth noting that determinate unfolding performed by CPD plays a huge role in these examples.
The default unfolding strategy implemented in \ecce allows for only a single non-determinate unfolding per a local control tree.
When considering conjunctions \lstinline{append$^o$ x y t /\ append$^o$ t z res} and \lstinline{max$^o$ x m /\ length$^o$ x l}, it unfolds the leftmost call which produces several branches in the tree.
The~rightmost call is only considered, if unfolding of its conjunction with the result of unfolding of the leftmost call produces only one result.
This indeed happens in these two examples.
If it was not to happen, then the conjunction would have been split at the global level and no deforestation or tupling would have been achieved.
It is not that hard to modify the examples in such a way so that CPD fails to transform them in a meaningful way.
For example, one extra disjunct can be added into the \lstinline{append$^o$} relation, or the calls to \lstinline{max$^o$} and \lstinline{length$^o$} may be reordered.
This is evidence of how non-trivial and fragile these transformers are.
More research should be done to make sure useful transformations are possible for many input programs.

\section{Conclusion}

In this paper we discussed some issues which arise in the area of partial deduction techniques for the relational programming language \mk.
We presented a novel approach to partial deduction --- conservative partial deduction --- which uses a heuristic to select a suitable relation call to unfold at each step of driving.
We compared this approach with the most sophisticated implementation of conjunctive partial deduction --- \ecce partial deduction system --- on 6 relations which solve 2 different problems.

Our specializer improved the execution time of all queries.
\ecce worsened the performance of all 4 implementations of the propositional evaluator relation, while improving the other queries.
Conservative partial deduction is more stable with regards to the order of relation calls than \ecce which is demonstrated by the similar performance of all 4 implementations of the evaluator of logic formulas.

Some queries to the same relation were improved more by ConsPD, while others --- by \ecce.
We conclude that there is still not one good technique which definitively speeds up every relational program.
More research is needed to develop models capable of predicting the performance of a relation which can be used in specialization of \mk.
There are some papers which estimate the efficiency of partial evaluation in the context of logic and functional logic programming languages~\cite{vidal2004cost,vidal2008trace}, and may facilitate achieving this goal.
Employing a combination of offline and online transformations as done in~\cite{hybrid} may also be the step towards more effective and predictable partial evaluation.
Other directions for future research include exploring how specialization influences the execution order of a \mk program, improving ConsPD so that it succeeds at deforestation and tupling more often, and coming up with a larger, more impressive, set of benchmarks.
\section*{Acknowledgements}
We gratefully acknowledge the anonymous referees and the participants of
VPT-2021 workshop for fruitful discussions and many useful suggestions.

\nocite{*}
\bibliographystyle{eptcs}
\bibliography{bibl.bib}

\end{document}